\documentclass[review]{elsarticle}
\newcommand{\caf}{CaF$_2$(Eu)}
\newcommand{\Ca}{$^{40}$Ca}
\newcommand{\F}{$^{19}$F}
\newcommand{\Cf}{$^{252}$Cf}
\newcommand{\Am}{$^{241}$Am}
\newcommand{\Cs}{$^{137}$Cs}
\usepackage{amssymb}
\usepackage{graphicx}	% required for `\includegraphics' (yatex added)

\begin{document}

\begin{frontmatter}
\title{Pulse-Shape Discrimination of \caf}
\date{\today}

\author[minowaken]{S. Oguri\fnref{gakushin}}
\ead{shugo@icepp.s.u-tokyo.ac.jp} 

\author[icepp]{Y. Inoue}

\author[minowaken]{M. Minowa}

\fntext[gakushin]{Research Fellow
 of the Japan Society for the Promotion of Science}

\address[minowaken]{Department of Physics, School of Science,
 University of Tokyo,
 7-3-1, Hongo, Bunkyou-ku, Tokyo 113-0033, Japan}
\address[icepp]{International Center for Elementary Particle Physics,
 University of Tokyo,
 7-3-1, Hongo, Bunkyou-ku, Tokyo 113-0033, Japan}

\begin{abstract}
 We measured the decay time of the scintillation pulses
 produced by electron and nuclear recoils in \caf\
 by a new fitting method.
 In the recoil energy region 5--30\,keVee,
 we found differences of the decay time
 between electron and nuclear recoil events.
 In the recoil energy region above 20\,keVee,
 we found that the decay time is independent
 of the recoil energy.
\end{abstract}


\end{frontmatter}

\section{Introduction}
\label{intro}

There is convincing evidence that most of the matter in our Galaxy
must be dark matter.
One of the most popular candidates for dark matter
is Weakly Interacting Massive Particles (WIMPs).
WIMPs are thought to be non-baryonic particles,
and the most plausible candidates for them are the Lightest
Supersymmetric Particles (LSPs) and the Lightest Kaluza-Klein Particles
(LKPs)\cite{Bertone2005279}.

WIMPs can be searched for directly,
as they interact with atomic nuclei in detectors.
Direct detection relies on one of two modes of interaction with target
nuclei.
The first mode is called spin-independent (SI) coupling.
It describes coherent interaction with the entire nuclear mass.
Therefore the SI part of the WIMP-nucleus cross section is large, if
the target nuclei have the large mass number.
For example, xenon and iodine are the favorable nuclei to detect WIMPs for
the SI interaction.
The second mode is called spin-dependent (SD) coupling.
It describes interaction of WIMP with the spin-content of the nucleus.
Hence the SD part of the WIMP-nucleus cross section is large, if the
spin-content of the target nuclei is large.
For example, \F\ is one of the most favorable nuclei to detect WIMPs for
the SD interaction because of its large nuclear spin.

Several direct WIMP searches using \F -based detectors,
such as bolometers\cite{Takeda2003145, Miuchi2003135},
a bubble chamber\cite{E.Behnke02152008},
scintillators\cite{Shimizu2006195,Belli199997,
Bernabei199773,Spooner199713,Bacci1994117},
a superheated droplet detector{\cite{archambault-2009}
and so on,
have already been performed.
In our group,
a WIMP search experiment using a \caf\ scintillator is carried out
at Kamioka observatory in 2005\cite{Shimizu2006195}.
\caf\ is one of the most suitable material among \F -based scintillator
because of its high light-output.

In other direct WIMP searches,
pulse shape discrimination (PSD) technique is used
in order to
statistically discriminate nuclear recoil signal events
from electron recoil background events.
A nuclear recoil event is different from an electron recoil event
in decay time of their pulse shapes.
In the PSD technique,
a distribution of the decay time constant is utilized
for the statistical discrimination.
The PSD feature in \caf\ was reported in the MeV region\cite{Bacci1994117,Belli200715}.
The pulse shape of \caf\ scintillator in lower energy region
relevant to the dark matter WIMP search
was measured previously by Tovey
{\itshape et al.} \cite{Tovey1998150}.
They reported that \caf\ has no PSD capability
in the 10--30 keVee region.
However, we recently measured the pulse shape of \caf\
and analysed our data by a more careful statistical method.
In this paper,
we report on the new measurement of \caf\ pulse shape and the analysis.
We found a difference in the pulse shape of \caf\ between
nuclear and electron recoil events.

\section{Experimental methods}

\subsection{Measurement system}

\begin{figure}
 \centering
 \includegraphics[width=10cm]{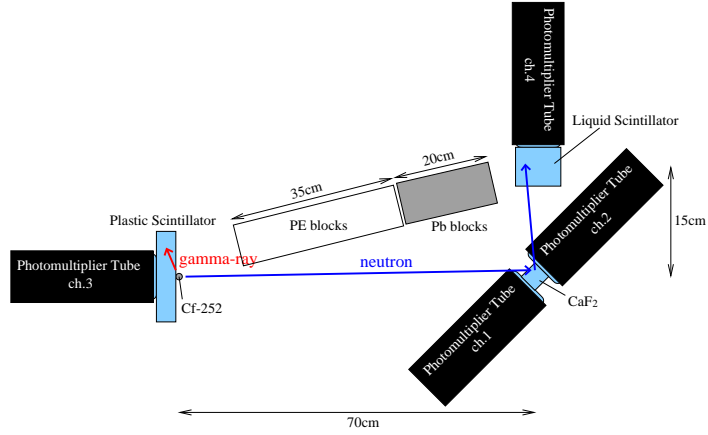}
 \caption{Schematic view of the neutron detectors.
 The neutron events is selected by
 the recoil energy in \caf\ scintillator and
 the TOF between the plastic scintillator and the liquid scintillator.}
 \label{souti_mosikizu}
\end{figure}

The scintillator sample consisted of
an unencapsulated, double-ended, 20-mm-diameter,
20-mm-long crystal of \caf\ with europium doping content of 0.5\%.
The crystal was glued on two 2-inch photomultiplier tubes
(Hamamatsu R329-02).

\Cf\ was used as a neutron source together with three detectors
as shown in Fig.\,\ref{souti_mosikizu}.
Nuclear recoil candidates
were selected by looking for coincidences between events in three
detectors.
The first one was the target \caf.
The second was a plastic scintillator,
which detected prompt $\gamma$ rays from fission decays of \Cf.
The last was a liquid scintillator (Saint-Gobain, BC-501A),
which caught scattered neutrons from the target \caf.

For the latter two detectors,
Hamamatsu R329-02 photomultiplier tubes were also used.
Signals from all the photomultiplier tubes
were sent to fast discriminators with sufficiently low thresholds.
A special care was taken to the two output signals of the \caf\ scintillator
so as to suppress multiple trailing pulses
due to its long decay time constant.
A coincidence of these two discriminator outputs were taken
to define the \caf\ scintillator hit
with a coincidence width of 160\,ns.
Coincidence widths of the plastic scintillator and \caf scintillator,
and the plastic scintillator and the liquid scintillator
were set to 195\,ns and 400\,ns, respectively.

The photomultiplier outputs
of the \caf\ and the liquid scintillators
and the discriminator output of the plastic scintillator
were recorded with a Tectronix TDS3034B digital oscilloscope
with sampling frequency of 2.5\,GHz and 300\,MHz bandwidth
by a trigger of the coincidence of three detectors.
The pulse shape of the \caf\ scintillator
and the timing of all the detectors
are thus recorded and sent to a computer.

The estimated solid angles of the \caf\ scintillator as seen
from the \Cf\ source
and of the liquid scintillator from the \caf\ scintillator
are $1.2 \times 10^{-3}$\,sr and $1.1 \times 10^{-1}$\,sr, respectively.
The scattering angle spanned by the three detectors was
about 90 degrees.
Between the plastic and liquid scintillators,
polyethylene (PE) blocks (35\,cm long)
and Pb blocks (20\,cm long)
are placed in order to shield the liquid scintillator
from neutrons coming directly from \Cf.

\subsection{Selection of nuclear recoils}

We distinguished nuclear recoil events from electron recoil events
using the time-of-flight (TOF) of the particle.
The time-of-flight of the neutron
from the \Cf\ source to the liquid scintillator is
\begin{equation}
 t_{{\rm SL}} = t_{{\rm ST}} + t_{{\rm TL}}
  = \frac{L_{{\rm ST}}}{\sqrt{2 E_{1} / m_{\rm n}}}
  + \frac{L_{{\rm TL}}}{\sqrt{2 E_{2} / m_{\rm n}}}.
\label{tof-theoretical-1}
\end{equation}
Here $t_{{\rm ST}}$ and $t_{{\rm TL}}$ are
the times-of-flight of the neutron
from the \Cf\ source to the target scintillator
and from the target scintillator to the liquid scintillator,
and $L_{\rm ST}$ and $L_{\rm TL}$ are the respective lengths.
In this measurement,
$L_{\rm ST}$ and $L_{\rm TL}$ are 70\,cm and 15\,cm.
$E_{1}$ and $E_{2}$ are the neutron energies
before and after the collision with the nucleus in the target,
and $m_{\rm n}$ is the neutron mass.
Then, the energy of the recoil nucleus is
\begin{equation}
 E_{\rm r} = E_{1} - E_{2}
  =  E_{1} \times \frac{2m_{\rm n}}{(M + m_{\rm n})^2} \times
  \left( M + m_{\rm n} \sin^2 \theta - \cos \theta \sqrt{
   M^2 - m_{\rm n}^2 \sin^2 \theta} \right)
\end{equation}
when the scattering is elastic.
Here $M$ is the recoil nucleus mass,
and $\theta$ is the neutron scattering angle.
In this measurement,
since the recoil angle $\theta$ is around 90 degrees,
the recoil energy is simply
\begin{equation}
 E_{\rm r} = E_{1} - E_{2}
  = E_{1} \times \frac{2 m_{\rm n}}{M + m_{\rm n}}.
\label{tof-theoretical-2}
\end{equation}
Inelastic cross section is negligibly small
in the neutron energy range\cite{ENDF} relevant to the present measurement
and therefore ignored in the analysis.

\begin{figure}
 \centering
 \includegraphics[width=10cm]{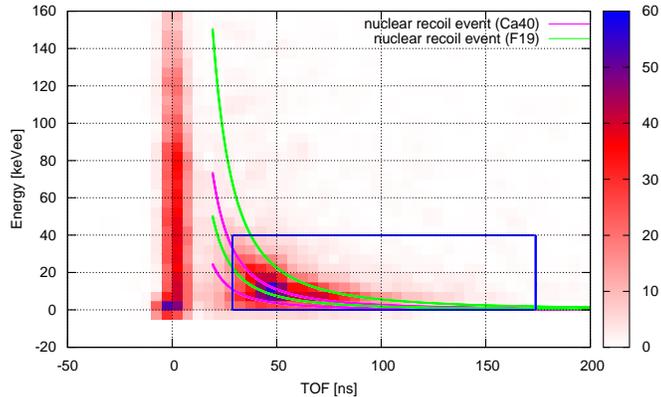}
 \caption{Scintillation energy and TOF distribution
 by density plot.
 The curves are the theoretical values
 of the scattering events
 for the \F\ and \Ca\ with quenching factors of 0.05 and 0.15.
 The inside of the rectangle is selected as neutron events.
 }
 \label{plot_tof}
\end{figure}

Density plots of the electron equivalent energy deposited in \caf\
versus TOF $t_{\rm SL}$
is shown in Fig.\,\ref{plot_tof}.
Theoretical expectation curves are also show in the Fig.\,\ref{plot_tof}
for the \F\ and \Ca\ with quenching factors of 0.05 and 0.15.
On the other hand,
Compton scattering events in the \caf\ scintillator caused
by gamma rays emitted by the \Cf\ source are expected to
gather in the region where TOF is 0\,ns.
Fig.\,\ref{plot_tof} shows clear separation of
the nuclear recoil events and the Compton scattering events.

To pickup the \F\ and \Ca\ recoil events,
we put a rectangle where nuclear recoil events are expected.
Consequently, we obtained enough pulse shapes for the PSD analysis.

\subsection{New method of PSD}
\label{PSD_method}

A pulse shape of scintillation is formed
by a train of small single-photoelectron pulses.
The arrival time relative to the pulse start time
of each pulse follows
an exponential distribution.
Hence, the scintillation pulse shape
has a characteristic time constant,
which is different between electron and
nuclear recoil events.

In many researches,
integrated pulse is used for analysis,
and an exponential rise function $A \times (1 - \exp[-(t-t_0)/\tau])$
is fitted to the data
by the least-squares method\cite{Smith1996299}.
This analysis is statistically not strictly correct
because integrated data points are not independent of each other.
Therefore, the data themselves before integration must be used
for the fitting analysis.

However, because the scintillation of \caf\ is not bright enough
and has large decay constant,
it is difficult to fit an exponential fall function
$A \times \exp[-(t-t_0)/\tau]$
to all the data points
by the least-squares method.
So we divided a train of data points of the oscilloscope
into 40\,ns bins,
and fitted the estimate number of photoelectrons
in each bin
calculated by an exponential fall function
to the data
by the maximum-likelihood method
via two parameters,
the estimate value of the first bin and the decay time constant.

The number of photoelectrons in each bin
follows a Poisson distribution.
So the probability distribution function $f(x_i, \mu_i)$
of the observed charge $x_i$
in $i$-th bin is
\begin{equation}
 f(x_i, \mu_i) = \sum_n
  \left[
   \mathrm{Poi}(n,\mu_i)
   \cdot \mathrm{Gau}(x_i, Q_0 + n Q_1, \sigma_0^2 + n \sigma_1^2)
  \right],
\end{equation}
where $\mu_i$ is the estimate value of the number
of photoelectrons in the $i$-th bin.
$\mathrm{Poi}(n,\mu)$ is a Poisson distribution function of $n$
with the mean value of $\mu$,
and $\mathrm{Gau}(x, Q, \sigma^2)$ is a Gaussian distribution of $x$
with the mean value of $Q$ and dispersion $\sigma^2$.
$Q_0$, $\sigma_0$, $Q_1$ and $\sigma_1$ are the parameters
for the conversion of the number of photoelectrons
to the charge recorded by the oscilloscope.
$Q_0$ and $\sigma_0$ are the pedestal and its width,
and $Q_1$ and $\sigma_1$ are one photoelectron response
and its standard deviation.
These four parameters are characteristic of PMTs
and they were measured by a single photoelectron spectra analysis
independently\cite{Tokar:683800}.
Tab.\,\ref{tab::QandSigma} shows the typical values of
$Q_0$, $\sigma_0$, $Q_1$ and $\sigma_1$
as measured in electron equivalent deposit energy.
The $Q_1$, $\sigma_0$ and $\sigma_1$ values are stable
within $\pm4.5\%$, $\pm16\%$ and $\pm23\%$ in measurement
after over a year, respectively.

\begin{table}[htbp]
 \centering
 \caption{Values of $Q_0$, $\sigma_0$, $Q_1$ and $\sigma_1$
 as measured in electron equivalent deposit energy.}
 \label{tab::QandSigma}
 \begin{tabular}{ccc}
  \hline \hline
  Parameter & PMT ch.1 [keV] & PMT ch.2 [keV] \\
  \hline
  $Q_0$      & $(1.1  \pm 0.3 ) \times 10^{-3}$
      & $(2.2  \pm 0.3 ) \times 10^{-3}$\\
  $\sigma_0$ & $(1.56 \pm 0.04) \times 10^{-2}$
      & $(1.83 \pm 0.04) \times 10^{-2}$\\
  $Q_1$      & $(3.15 \pm 0.07) \times 10^{-1}$
      & $(3.06 \pm 0.07) \times 10^{-1}$\\
  $\sigma_1$ & $(1.70 \pm 0.08) \times 10^{-1}$
      & $(1.69 \pm 0.08) \times 10^{-1}$\\
  \hline
 \end{tabular}
\end{table}

From this, the likelihood function $L(\mu_0, \tau)$ is
\begin{equation}
 L(\mu_0, \tau) = \prod_{\rm PMT\,ch.1,2}
  \prod_i f \left(x_i, \mu_0 \cdot
  \exp\left(- \frac{t \cdot (i-1)}{\tau}\right) \right),
\end{equation}
where $t$ is the bin width.
The two parameters, $\mu_0$ (the estimate value of the photoelectron
number in the first bin) and $\tau$ (the decay time constant),
are evaluated when they maximize the likelihood function
$L(\mu_0, \tau)$.

\section{Result and discussion}

\begin{figure}
 \centering
 \includegraphics[width=10cm]{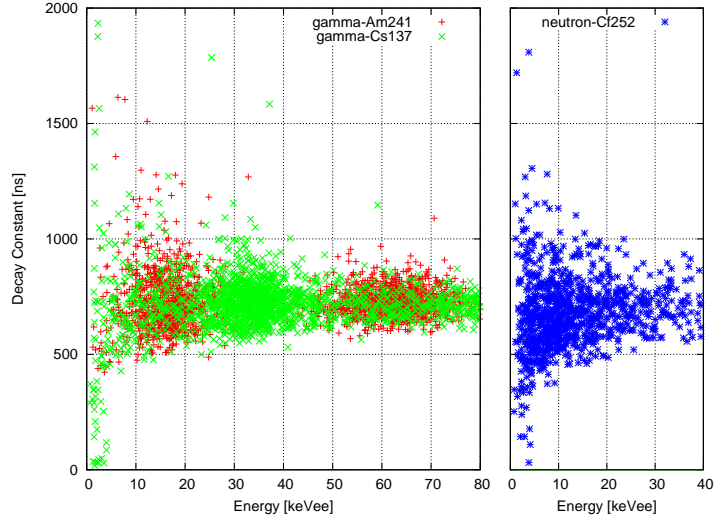}
 \caption{The scatter plot of the experimental data.
 The decay constant is evaluated by the maximum likelihood method.
 (Sec.\,\ref{PSD_method})}
 \label{sampu_kekka}
\end{figure}

A sample of the result of the maximum likelihood fitting
to the experimental data is shown in Fig.\,\ref{sampu_kekka}
for $\gamma$ ray events with \Am\ and \Cs,
and for selected neutron events with \Cf.
A difference is seen in the decay time distribution
between electron and nuclear recoil events.

\begin{figure}
 \centering
 \includegraphics[width=10cm]{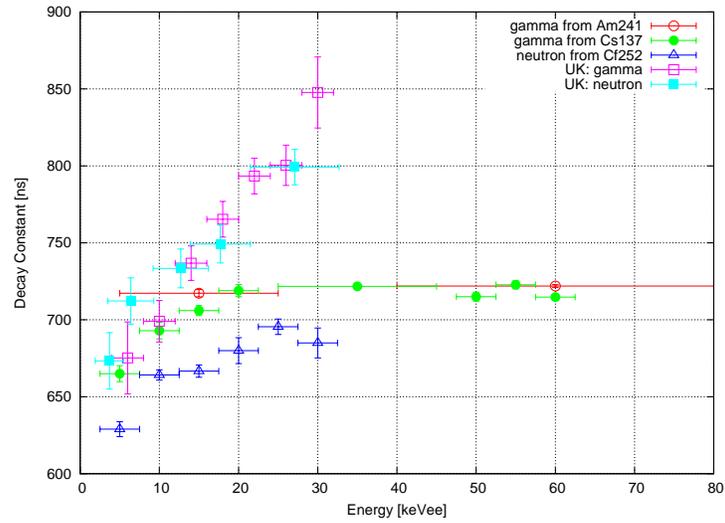}
 \caption{The decay time constant versus the energy.
 The present result is shown by open circle,
 filled circle and open triangle signs.
 The results in Ref.\,\cite{Tovey1998150} are open and filled squares.}
 \label{plot_kekka}
\end{figure}

In order to see the difference more clearly,
we divided the events
into electron equivalent energy bins of 5\,keV width,
and evaluated the mean value of the decay constant
in each bin.
Exceptionally,
we put regions of photoelectric absorption peaks,
13.9\,keVee (\Am), 59.5\,keVee (\Am) and 32.2--36.4\,keVee (\Cs),
into wider single bins.
The distribution of the decay time constant $\tau$
can be approximated by a Gaussian in $\ln(\tau)
$\cite{Kudryavtsev1999167}:
\begin{equation}
 \frac{dN}{d\tau} =
  \frac{N_{_0}}{\tau \sqrt{2\pi} \ln w}
  \cdot \exp \left[
	      \frac{-(\ln \tau - \ln \tau_{_0})^2}{2(\ln w)^2}
	     \right].
\end{equation}
We evaluate three parameters,
$\tau_{_0}$ (the mean value of the exponent),
$N_{_0}$ (the number of events) and
$w$ (a width parameter),
in each bin by the least-squares method.
Fig.\,\ref{plot_kekka} shows the evaluated mean decay time constants $\tau_{_0}$
as a function of energy.
There are marked differences of the decay time
between electron and nuclear recoil events.

The results in Ref.\,\cite{Tovey1998150} are
also shown in Fig.\,\ref{plot_kekka}.
In their results,
the differences are not seen.
Moreover,
the decay constants rise rapidly with energy
from $\sim$\,660\,ns to $\sim$\,850\,ns.
In the present work,
a similar dependence exists in the low energy region,
but it is weak.

Strictly speaking,
because the pulse shape is not a single exponential curve,
but the sum of a few exponential curves,
the value of the decay time constant
is dependent on the way of the analysis.
So,
the values of different measurements
cannot be compared directly.
However,
these properties,
the difference between the nuclear recoil and the electron recoil
and the relation between the decay constant and the recoil energy,
are independent of the way of the analysis.

\section{Conclusion}

We developed a new method
to evaluate the decay time constant of scintillation
and applied it to the \caf\ scintillator.
We found the difference in the scintillation decay constants
between electron and nuclear recoil events
(Fig.\,\ref{plot_kekka}).
We also concluded that
the decay time constant of \caf\ is constant
in the recoil energy region above 20\,keVee.

\bibliographystyle{elsarticle-num}
\bibliography{bunken.bib}

\end{document}